\documentclass[aps,pra,superscriptaddress,twocolumn,showpacs,longbibliography,titlepage=false]{revtex4-1}
\usepackage{graphicx}
\usepackage{amsmath}
\usepackage{amssymb}
\usepackage{subfigure}
\usepackage{natbib}
\usepackage{color}
\usepackage{blkarray}
\usepackage{dsfont}
\usepackage{xspace}
\usepackage[breaklinks=true,colorlinks=true,linkcolor=blue,urlcolor=blue,citecolor=blue]{hyperref}
\usepackage[percent]{overpic}
\usepackage{mathrsfs}
\usepackage{verbatim}

\def\ket#1{{\left| #1 \right\rangle}}

\newcommand{\cnot}{\ensuremath{\mathsf{CNOT}}\xspace}
\newcommand{\pvalueword}{p-value\xspace}
\newcommand{\pvaluesword}{p-values\xspace}
\newcommand{\pvalue}{p}
\newcommand{\prob}{\mathsf{p}}
\newcommand{\probvec}{\boldsymbol{\mathsf{p}}}
\newcommand{\counts}{\mathsf{x}}
\newcommand{\countsvec}{\boldsymbol{\mathsf{x}}}
\newcommand{\fulldataset}{x}

\begin{document}
\title{Probing context-dependent errors in quantum processors}

\author{Kenneth Rudinger}
\affiliation{Center for Computing Research, Sandia National Laboratories, Albuquerque, NM 87185, USA}
\email{kmrudin@sandia.gov} 
\author{Timothy Proctor}
\affiliation{Sandia National Laboratories, Livermore, CA 94550, USA}
\author{Dylan Langharst}
\affiliation{Center for Computing Research, Sandia National Laboratories, Albuquerque, NM 87185, USA}
\affiliation{Behrend College, Pennsylvania State University, Behrend, Erie, PA 16563}
\author{Mohan Sarovar}
\affiliation{Sandia National Laboratories, Livermore, CA 94550, USA}
\author{Kevin Young}
\affiliation{Sandia National Laboratories, Livermore, CA 94550, USA}
\author{Robin Blume-Kohout}
\affiliation{Center for Computing Research, Sandia National Laboratories, Albuquerque, NM 87185, USA}
\date{\today}

% ========================= Abstract ===========================%
\begin{abstract} \noindent
Gates in error-prone quantum information processors are often modeled using sets of one- and two-qubit process matrices, the standard model of quantum errors.  However, the results of quantum circuits on real processors often depend on additional external ``context'' variables.  Such contexts may include the state of a spectator qubit, the time of data collection, or the temperature of control electronics.  In this article we demonstrate a suite of simple, widely applicable, and statistically rigorous methods for detecting context dependence in quantum circuit experiments.  They can be used on any data that comprise two or more ``pools'' of measurement results obtained by repeating the same set of quantum circuits in different contexts.  These tools may be integrated seamlessly into standard quantum device characterization techniques, like randomized benchmarking or tomography. We experimentally demonstrate these methods by detecting and quantifying crosstalk and drift on the publicly accessible 16-qubit ibmqx3.
\end{abstract}
\maketitle

\section{Introduction}
\noindent Quantum characterization, verification, and validation (QCVV) \cite{blume2016certifying,merkel2013self, greenbaum2015introduction, dehollain2016optimization, knill2008randomized, magesan2011scalable, gambetta2012characterization, proctor2018direct, mckay2017three, francca2018approximate, magesan2012efficient, sheldon2016characterizing, chasseur2017hybrid, wood2017quantification, carignan2015characterizing, hashagen2018real, brown2018randomized, emerson2005scalable, emerson2007symmetrized, knill2008randomized, wallman2015robust, wallman2015estimating} tools provide a ways to probe the \emph{in situ} behavior of quantum information processing hardware.  Most QCVV protocols assume a ``standard model'' of errors in which each imperfect quantum operation is represented by a single, completely positive, trace preserving (CPTP) linear map on density matrices (i.e., a \emph{process matrix}).  Although this model can describe many deviations from ideal behavior, including coherent errors caused by a fixed Hamiltonian and stochastic errors caused by white noise fluctuations, there are many other possible failure modes whose impacts on both quantum error correction (QEC) and near-term quantum information processing applications are not yet well understood.  Many of them manifest as \emph{dependence} of the error process on some external variable, or \emph{context}, that isn't supposed to affect qubit behavior \cite{Veitia2017}.  For example, an error rate might drift over time \cite{fogarty2015nonexponential,Grace2012,dehollain2016optimization,van2013quantum}, or increase when a nearby qubit is being measured or driven \cite{gambetta2012characterization,mckay2017three,proctor2018direct,Plitz2014,Rigetti2010,Altomare2010}.  These effects are important in their own right. They might contribute significantly to the device's total observed error rate \cite{gambetta2012characterization,mckay2017three,proctor2018direct}, and they may have consequences for QEC \cite{Baireuther2018machinelearning,Jin2012,Florjanczyk2016,Greenbaum2018,Plitz2014,Buterakos2018}.  Context dependence is also important because it can interfere with standard QCVV techniques such as randomized benchmarking (RB) \cite{knill2008randomized, magesan2011scalable, gambetta2012characterization, proctor2018direct,francca2018approximate, mckay2017three, magesan2012efficient, sheldon2016characterizing, chasseur2017hybrid, wood2017quantification, carignan2015characterizing, hashagen2018real, brown2018randomized, emerson2005scalable, emerson2007symmetrized, knill2008randomized, wallman2015robust, wallman2015estimating}
 or gate-set tomography (GST) \cite{blume2016certifying,merkel2013self,greenbaum2015introduction,dehollain2016optimization}, and potentially invalidate conclusions drawn from them \cite{van2013quantum}.

\begin{figure}
\includegraphics[width=1\columnwidth]{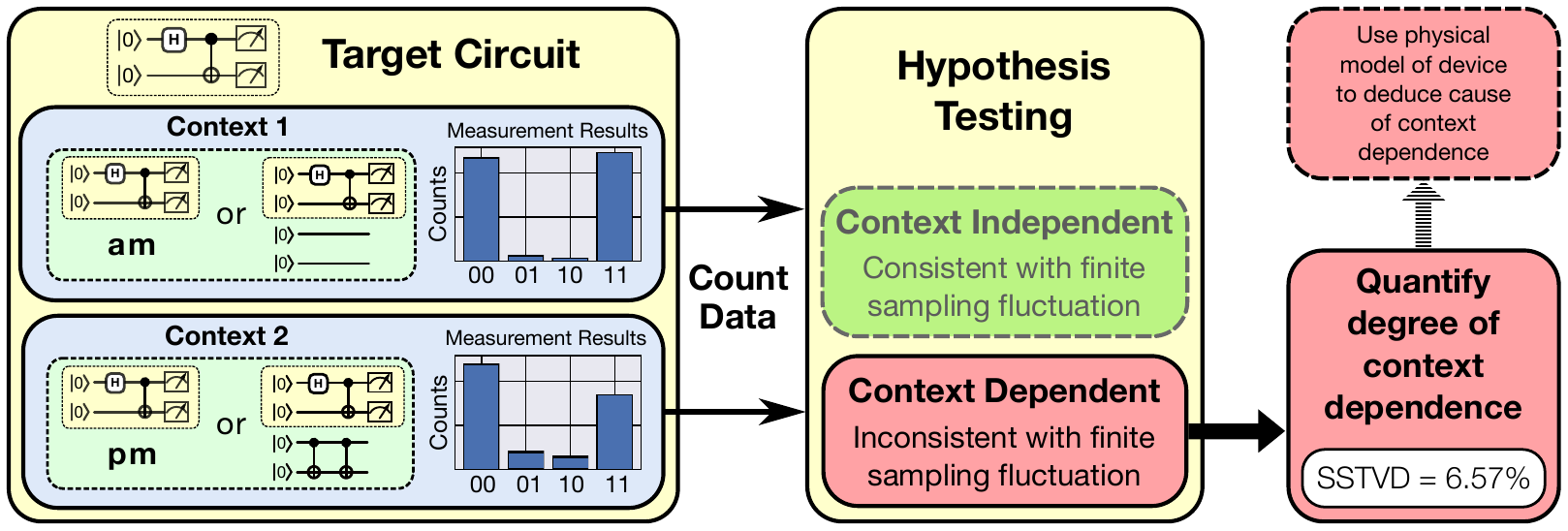}
\caption{\label{fig:schematic} An illustration of how to detect and quantify context dependence in a quantum information processor by repeatedly performing a quantum circuit in two or more contexts. In this simple example, a Bell state is prepared during two different time periods (am/pm), to test for time variation; or while an adjacent pair of qubits is or is not being driven, to test for crosstalk. The measurement outcome frequencies for the two contexts are compared to determine if the circuit behavior is the same across contexts. If not, the change is quantified. Multiple test circuits and a physical model of the device can sometimes enable identification of the underlying cause and indicate the size of the effect.}
\end{figure}

In this paper we propose and demonstrate a practical, statistically rigorous toolkit for detecting whether a quantum circuit's observable behavior depends on external variables. The underlying statistical tasks here are old and well studied \cite{wasserman2013all,lehmann2006testing,wilks1938large,agresti2012}, so we make no claims of statistical novelty. Instead, our focus is on choosing and harnessing established statistical techniques for detecting context dependence in QCVV, using the type of data most often found in quantum device characterization and circuit-based experiments. Almost all such experiments generate \emph{count data}: the aggregated outcomes of $N$ repetitions of one or more quantum circuits that each begin with a state preparation and end with a measurement.

Usually, all the measurement results for a single circuit are collected into a single ``pool''. This precludes testing for variation, because a single pool of counts is always perfectly consistent with a single underlying set of probabilities for the observed outcomes.  However, some data have additional structure, such as time stamps, that define a natural division into two or more pools that are each associated with a different ``context''.  Then, we can look for \emph{significant} variation in the circuit behavior between contexts (Fig.~\ref{fig:schematic}).  For example, flipping two coins 100 times and getting 49 heads for one coin and 55 for the other is intuitively consistent with the claim that the coins are identically biased; the variation is typical of random finite-sample fluctuations. Observing instead 28 heads for one coin and 72 heads for the other is strong evidence that the coins actually have different biases. We can address this question formally using \emph{statistical hypothesis testing}, a standard framework for rigorously deciding if there is sufficient evidence to reject a base assumption, known as a \emph{null hypothesis}. In the tools we propose, our null hypothesis is that there is no context dependence, and we seek statistically significant evidence in the data to the contrary.

This paper is structured as follows. In Section~\ref{sec:detection} we present hypothesis testing techniques for detecting context dependence in count data from one or more circuits. In Section~\ref{sec:quantification} we adapt these context dependence \emph{detection} tools to the task of context dependence \emph{quantification}. In Section~\ref{sec:sim} we simulate applying these techniques to detect drift, demonstrating that these methods can clearly highlight context-dependent errors. In Section~\ref{sec:exp} we apply our techniques to drift and crosstalk detection and quantification on the ibmqx3 \cite{ibmqx}, a publicly accessible superconducting quantum processor. In Section~\ref{sec:srb} we discuss the relationship between our tools and simultaneous RB \cite{gambetta2012characterization}, a popular crosstalk quantification technique, and we conclude in Section~\ref{sec:conclusions}.

\section{Detecting context dependence}\label{sec:detection}
\subsection{Single circuit data}
First, we consider how to \emph{detect} context dependence in a \emph{single} quantum circuit.  Suppose this circuit has $M\geq2$ possible measurement outcomes, indexed by $m=1,2,\dots,M$.  In general, if a circuit has $n$ qubits (and all $n$ qubits are read out at the end of the circuit), then $M=2^n$.  Note that we could also choose to measure only a subset of the qubits in the system, or marginalize multi-qubit data over some of the qubits. Let this circuit be performed repeatedly in each of $C$ different contexts, indexed $c=1,2,\dots,C$. For example, the contexts might correspond to  distinct time intervals, or to driving (or not driving) neighboring qubits (see Fig.~\ref{fig:schematic}). For each context $c$, the circuit defines a probability distribution over the possible measurement results
\begin{equation}
\probvec_c = (\prob_{c,1},\prob_{c,2},\ldots,\prob_{c,M}).
\label{eq:pc}
\end{equation}
These are probabilities for obtaining each of the $M$ measurement outcomes, \emph{after} averaging over any other unaccounted-for contexts that might vary within a $c$-indexed context. For example, time is a continuously varying context variable, and a time period context is a coarse-graining over time.  Thus, in this example each $\probvec_c$ is the probability distribution after this time-averaging.
An experiment consists of running our circuit $N_c$ times in each context $c$ and recording the total counts for each measurement outcome $m$.  This effectively samples from each of the the $\probvec_c$ distributions, producing measurement results $\fulldataset=\{\countsvec_c\}$. Here
\begin{equation}
\countsvec_c = (\counts_{c,1},\counts_{c,2},\ldots,\counts_{c,M}),
\label{eq:xc}
\end{equation}
 is a vector of positive integers summing to $N_c$, representing the observed counts from $N_c$ repeats of the circuit in context $c$. In terms of the data, context independence holds iff all of the data were drawn from the same underlying probability distribution $\probvec_0$.  To detect context dependence we therefore ask whether the measurement results in different contexts are consistent with being drawn from a single distribution. This is a hypothesis testing problem: we are looking for evidence to reject the \emph{null hypothesis} that the underlying distributions are context independent.

In general, hypothesis testing is the following procedure:
\begin{enumerate}
    \item Choose a \emph{statistic}. This is a function $\Lambda$ from the space of all possible experimental results to $\mathbb{R}$.
    \item Choose a significance threshold level $\alpha \in (0,1)$. A popular choice is $\alpha = 5\%$, corresponding to a $95\%$ confidence.
    \item Collect data ($\fulldataset$) and evaluate $\Lambda(\fulldataset)$.
    \item Calculate the \pvalueword ($\pvalue$) of $\Lambda(\fulldataset)$. This is the probability of observing a value of $\Lambda$ that is at least as extreme as $\Lambda(\fulldataset)$ \textbf{if} the null hypothesis is true.
    \item Reject the null hypothesis if $\pvalue< \alpha$. Here, rejecting the null hypothesis means detecting context dependence.
\end{enumerate}
Any procedure of this form ensures that the probability of falsely detecting context dependence is at most $\alpha$.  Within this constraint, it is desirable to choose a procedure -- i.e., a statistic -- with high \emph{power} to detect context dependence if it is present. For general hypothesis testing, there is no universally optimal statistic except for the simplest problems \cite{lehmann2006testing}, but the log-likelihood ratio (LLR) statistic is canonical and popular, and we have found it to be convenient and powerful.

For data $\fulldataset$, a statistical model parameterized by $\theta \in \mathcal{H}$ for some parameter space $\mathcal{H}$, and a null-hypothesis subspace $\mathcal{H}_0 \subset \mathcal{H}$, the LLR is defined as
\begin{equation}
\lambda  :=  - 2 \log [\mathcal{L}(\hat{\theta}_{0})/\mathcal{L}(\hat{\theta})],
\end{equation}
 where  $\mathcal{L}(\theta) = \text{Pr}(\theta \,|\, \fulldataset)$ is the likelihood function, $\hat{\theta}_{0}$ is the maximum likelihood estimate of $\theta$ over the null-hypothesis subspace $\mathcal{H}_0$, and $\hat{\theta}$ is the maximum likelihood estimate of $\theta$ over the full parameter space $\mathcal{H}$ \cite{wasserman2013all,lehmann2006testing,wilks1938large}. For our problem, we have
 \begin{enumerate}
     \item $\mathcal{H}_0$: the null hypothesis that $\probvec_{c}=\probvec_0$ for all $c$. The maximum likelihood estimate over the null hypothesis space is 
     $ \hat\probvec_0 = N^{-1}(\counts_1,\counts_2,\ldots,\counts_M)$, with $\counts_m = \sum_c \counts_{c,m}$ counts obtained by aggregating over contexts, and $N = \sum_c N_c$.
     \item $\mathcal{H}$: the alternative hypothesis that each $\probvec_c$ is independent. The maximum likelihood estimate under the alternative hypothesis is
      $\hat\probvec_c = {\countsvec_{c}/N_c}$.
 \end{enumerate}
Via basic multinomial statistics, the LLR is then
\begin{equation}
\lambda =-2 \sum_{m=1}^{M} \left[ \counts_{m} \log\left( \frac{\counts_{m}}{N} \right) - \sum_{c=1}^{C} \counts_{c,m} \log\left(\frac{ \counts_{c,m}}{N_c}\right)\right].
\label{eq:llr}
\end{equation}

To compute \pvaluesword, we appeal to Wilks' theorem \cite{wilks1938large}. It states that if the null hypothesis holds, as the number of samples $\to \infty$, the LLR converges to a $\chi_k^2$ random variable, where $k= l - l_0$ and $l$ (resp., $l_0$) is the number of free parameters in the full (resp., null) model \cite{wasserman2013all,lehmann2006testing,wilks1938large}. Each probability vector contains $M-1$ free parameters ($M$ probabilities summing to 1), so $l = C(M-1)$ and $l_0 = (M-1)$.  If $N_c \gg 1$, then under the null hypothesis $\lambda$ is approximately $\chi^2_{k}$ distributed, with
\begin{equation}
k= (C - 1)(M - 1).
\label{eq:q-dof}
\end{equation}
 The \pvalueword of an observed $\lambda$ is therefore approximated by
\begin{equation}
\pvalue \approx 1-F_k(\lambda),
\label{eq:pvalue}
\end{equation}
 where $F_k$ is the $\chi^2_{k}$ cumulative distribution function. For pre-specified $\alpha$, we say that context dependence has been detected at significance $\alpha$ if $\pvalue<\alpha$. We call this simple primitive the \emph{individual circuit test} (ICT), because it applies to data from a single circuit.

Here is a simple example of how the ICT can be used to detect context dependence.  Consider a 1-qubit circuit comprising preparation of $\ket{0}$, application of $X_{\pi/2} = \exp(-i\pi\sigma_x/4)$, and measurement of $\sigma_z$.  It is performed in two contexts: (1) while a neighbor qubit sits idle; (2) while the neighbor is driven in some fashion.  Now, suppose the operations are perfect under Context 1, but the driving in Context 2 causes the $X_{\pi/2}$ gate to over-rotate: $X_{\pi/2} \to \exp(-i1.1\pi\sigma_x/4)$. We chose a significance level of 5\%, and simulated $200$ repetitions of the circuit in each context, observing 99 ``0'' outcomes in Context 1 and 131 in Context 2. Putting this data into Eqs.~(\ref{eq:llr} -- \ref{eq:pvalue}) with $C=2$ and $M=2$, we find that the \pvalueword is $\pvalue\approx0.1\%$. This is easily significant at the $5\%$ level ($\pvalue < 5\%$), so context dependence was detected in this simulated experiment. We also simulated a scenario where driving did \emph{not} cause any change, and this time obtained 108 ``0'' counts in Context 1 and 107 in Context 2. Calculated in the same way, the \pvalueword for this data was $\pvalue \approx 92\%$, so context independence was not rejected. If we repeated this simulation many times, in the latter case where there is no context dependence we'd expect to erroneously detect context dependence in 5\% of the trials.

\subsection{Multi-circuit data}
Many quantum circuit based experiments involve collecting data from multiple distinct circuits, as is the case for most QCVV techniques, including all RB protocols \cite{knill2008randomized, magesan2011scalable, gambetta2012characterization, proctor2018direct,francca2018approximate, mckay2017three, magesan2012efficient, sheldon2016characterizing, chasseur2017hybrid, wood2017quantification, carignan2015characterizing, hashagen2018real, brown2018randomized, emerson2005scalable, emerson2007symmetrized, knill2008randomized, wallman2015robust, wallman2015estimating}, GST \cite{blume2016certifying,merkel2013self,greenbaum2015introduction,dehollain2016optimization} and other tomographic methods \cite{kimmel2015robust,rudinger2017experimental}.  We now extend the context dependence detection method presented above to the multi-circuit scenario. Consider $Q$ circuits indexed $q=1,2,\dots,Q$, each with $M$ possible outcomes, indexed $m=1,2,\dots M$ \footnote{The generalization to $q$-dependent $M$ is avoided mostly only for notational simplicity.}. These circuits are all implemented in each of $C$ contexts, again indexed by $c$ for $c=1,2,\dots,C$. Slightly generalizing the notation of Eq.~\eqref{eq:pc}, let
\begin{equation}
\probvec_{q,c} = (\prob_{q,c,1},\prob_{q,c,2},\ldots,\prob_{q,c,M}),
\end{equation}
denote the underlying probability distribution for circuit $q$ in context $c$. As before, a particular circuit is context independent iff all $\probvec_{q,c}=\probvec_{q,0}$ for some circuit-dependent $\probvec_{q,0}$. All of the circuits are context independent if this holds for all circuits $q$.

Consider data generated by $N_{q,c}$ repeats of circuit $q$ in context $c$. Let $\counts_{q,c,m}$ denote counts data for outcome $m$ of circuit $q$ in context $c$, with the full set of data denoted by
\begin{equation}
\fulldataset = \{ \countsvec_{q,c}=(\counts_{q,c,1},\counts_{q,c,2},\dots,\counts_{q,c,M})\}.
\end{equation}
There are many ways to test for context dependence with multi-circuit data of this sort. Most obviously, we could apply the ICT defined above to the data from each circuit, to separately test for context dependence in each circuit. However, implementing all $Q$ ICTs involves implementing multiple statistical hypothesis tests, and it is necessary to take this into account. If the null hypothesis is true, and we naively implement $T$ independent hypothesis tests all at some fixed significance $\alpha$, then we expect approximately $\alpha T$ of the tests to falsely reject the null hypothesis just by random chance. In fact, the probability of falsely rejecting the null hypothesis in at least one test will converge to 1 as $T$ increases.

To keep the probability of false detection in one or more tests -- known as the \emph{family-wise error rate} (FWER) \cite{lehmann2006testing,shaffer1995multiple} --  to at most $\alpha$, it is necessary to adjust the significance of the individual tests. The simplest solution is the \emph{generalized Bonferroni correction} \cite{lehmann2006testing,shaffer1995multiple}:  For any tests implemented together, a FWER of at most $\alpha$ can be obtained by setting the ``local" significance level of test $i$ to $\alpha_i = \alpha w_i$ for any $w_i\geq0$ satisfying $\sum_iw_i=1$.  Implementing all $Q$ ICTs with each significance set to $\alpha/Q$ is therefore sufficient to maintain a global significance of $\alpha$. However, the Bonferroni correction is unnecessarily conservative, so we will use a strictly more powerful correction.

Because the $\lambda_q$ are independent under the null hypothesis, where $\lambda_q$ is the LLR for circuit $q$, we can implement the ICTs with a \emph{Hochberg correction} \cite{hochberg1988sharper,shaffer1995multiple}\footnote{More powerful corrections are also possible}. In this setting, the Hochberg correction keeps the FWER to at most $\alpha$ using the following procedure:
\begin{enumerate}
\item Order the $Q$ \pvaluesword from smallest to largest: $\pvalue_{(1)}$, $\pvalue_{(2)},$ $\dots,$ $\pvalue_{(Q)}$.
\item Find the largest $l$ such that $\pvalue_{(l)} \leq \alpha/(Q-l+1)$, denoting this integer by $l_{\max}$.
\item Reject the null hypothesis (context independence) for all circuits with \pvaluesword smaller than
\begin{equation}
\pvalue_{\text{threshold}} = \alpha/(Q-l_{\max}+1).
\label{pthreshold}
\end{equation}
\end{enumerate}
Hereafter, we use this multi-test correction procedure used for the ICTs herein. Note that $\pvalue_{\text{threshold}}$ is not a true threshold for the statistical significance of a \pvalueword, in the sense that it depends on the data.  We therefore refer to it instead as a ``pseudo-threshold". Sometimes it is convenient to convert this to a pseudo-threshold above which the LLR of a circuit is significant. Inverting Eq.~\eqref{eq:pvalue}, this is given by
\begin{equation}
\lambda_{\text{threshold}} = F^{-1}_{k}(1-\pvalue_{\text{threshold}}),
\label{eq:llrthreshold}
\end{equation}
where $k$ is the degrees of freedom per circuit, in Eq.~\eqref{eq:q-dof}, and $F^{-1}_k$ is the inverse cumulative distribution function for the $\chi^2_k$ distribution.  

The ICTs are often not the most sensitive for deciding whether there is context dependence in at least one circuit. In particular, there are tests that are more sensitive to context dependence that is distributed uniformly over all the circuits. A complementary test statistic, powerful for detecting uniformly distributed context dependence, is the \emph{aggregate} LLR
\begin{equation}
\lambda_\text{agg}=\sum_{q=1}^Q\lambda_{q},
\end{equation}
 where, again, $\lambda_q$ is the LLR for circuit q. This is the LLR between the null hypothesis of context independence in \emph{all} circuits and the full context dependence model. That is, it is the LLR between the model whereby $\probvec_{q,c}=\probvec_{q,0}$ for some $\probvec_{q,0}$ and all $q$, and the model whereby all the $\probvec_{q,c}$ are independent. Therefore, when the null hypothesis holds, $\lambda_\text{agg}$ approximately follows a $\chi^2_{k_{\text{agg}}}$ distribution with
\begin{equation}
k_{\text{agg}}=Q(C-1)(M-1).
\end{equation}

For $k \gg 1$, the $\chi^2_{k}$ distribution is approximately normal with mean $k$ and variance $1/(2k)$. Therefore, in the common situation of $Q \gg 1$, a convenient and intuitive way to express the statistical significance of $\lambda_\text{agg}$ is as the number of standard deviations by which it exceeds its expected context-\emph{independent} value. This is given by
\begin{equation}
\mathcal{N}_\sigma = \frac{\lambda_{\text{agg}}-k_{\text{agg}}}{\sqrt{2k_{\text{agg}}}}.
\label{eq:nsigma}
\end{equation}
In our experience, the \pvalueword of the aggregate LLR is often vanishingly small (see, e.g., Sec.~\ref{sec:sim}), so $\mathcal{N}_\sigma$ provides an alternative measure of statistical significance that is on a more convenient scale. It is sometimes useful to have a threshold for $\alpha$ significance of the $\mathcal{N}_\sigma$, and this is given by
\begin{equation}
\mathcal{N}_{\sigma,\text{threshold}} = \frac{ F^{-1}_{k_{\text{agg}}}(1-\alpha)-k_{\text{agg}}}{\sqrt{2k_{\text{agg}}}}.
\label{eq:nsigmathreshold}
\end{equation}
When $Q \gg 1$, this is essentially identical to the standard significance thresholds for standard deviations above the mean with a normal distribution.

Although the aggregate LLR test is often more sensitive, the ICTs are useful because they indicate \emph{which} circuits vary. This can constitute helpful diagnostic information, as demonstrated later. We can strike a balance between these tests by implementing the set of ICTs \emph{and} the aggregate test, with significance levels adjusted appropriately. A reasonable strategy, which we adopt for the simulations and experiments in this paper, is the following. For a user-specified global significance $\alpha$:
\begin{enumerate}
\item Implement the aggregate test at significance level $\alpha/2$. If context dependence is detected set $\beta = \alpha$; otherwise set $\beta = \alpha/2$.
\item Implement the ICTs using a Hochberg correction at a significance of $\beta$.
\end{enumerate}
This type of multi-test compensation is based on the \emph{closed test principle} (a generalization of the Bonferroni correction), and it controls the FWER to be at most $\alpha$~\cite{bretz2009graphical}.

\subsection{Choosing the circuits}
The context dependence detection methods that we have proposed in this section can be applied to data from almost any set of circuits.  They can be bolted on to almost any device characterization protocol. However, if context dependence detection is a high priority, it is often useful to choose circuits that are sensitive to all the parameters that might vary with context. GST circuits \cite{blume2016certifying,merkel2013self,greenbaum2015introduction,dehollain2016optimization} are one reasonable choice, because they are informationally complete for tomography of gates, state preparations and measurements (SPAM). If context dependence manifests as an observable dependence of gate or SPAM process matrices on the context, at least one GST circuit will be sensitive to it. We use GST circuits in our examples below.

Using our tools on data from GST circuits does \emph{not} require implementing the tomographic reconstructions of GST.  Tomographic reconstructions using the data from each context are nevertheless clearly possible with GST data.  This naturally raises the question of what our tools add that couldn't be achieved as easily with tomography. Our tools have three distinct advantages over tomography, which highlight how they complement any tomographic data analysis. First, precise tomography require large amounts of data and many individual circuits, whereas detecting context dependence can often be achieved using few circuits and/or less data. Second, tomographic methods are based on fitting a model, and become unreliable if this model does not accurately describe the system \cite{van2013quantum}. In contrast, these direct context dependence detection tools require no model of the underlying operations (the gates and SPAM). Finally, tomography is computationally expensive, but the tools here require only very simple classical computation.

\section{Quantifying context dependence}\label{sec:quantification}
The detection methods presented in the previous section \emph{test} whether or not there is statistically significant evidence of context dependence; when used rigorously they only report ``yes'' or ``no''. In general, the value of a test statistic will not necessarily quantify the ``strength'' of a detected effect. Neither the magnitude of the LLR for each circuit, nor the aggregate LLR, nor the associated \pvaluesword, nor the aggregate $\mathcal{N}_{\sigma}$ directly quantify the strength of context dependence. Instead, they quantify our \emph{confidence} that context dependence exists.  If there is \emph{any} context dependence in one or more circuits then, as we take more data, both $\lambda_{\text{agg}}$ and $\mathcal{N}_{\sigma}$ will increase without bound.  Arguably, the most interesting metrics of context dependence ``strength'' would describe the variation of an underlying gate/SPAM error rate, but this is the domain of specific QCVV protocols (e.g. RB or GST).  In the very general framework of this paper, the most we can do is to quantify the strength of each individual circuit's context dependence. This is equivalent to estimating how much the circuit's outcome probabilities change between contexts, and there are many ways to do this.
\subsection{Jensen-Shannon Divergence}\label{sec:jsd}
The simplest way to quantify context dependence is to rescale the per-circuit LLRs to
\begin{equation}
\text{JSD}_{q}  = \frac{\lambda_\text{q}}{2N_q},
\label{eq:JSDq}
\end{equation}
where $N_q = \sum_{c} N_{q,c}$. As suggested by this notation, $\text{JSD}_{q}$ provides an estimate of the Jensen-Shannon divergence (JSD) of the underlying probability distributions. For probability distributions $P_c$ over $M$ events, with $c = 1,2,\dots,C$, and some weightings $\pi_c$ with $\sum_c \pi_c=1$, the JSD is defined by \cite{lin1991divergence}
\begin{equation*}
\text{JSD}_{\{\pi_c\}}(P_1,\dots,P_C) = H\left(\sum_{c=1}^C \pi_cP_c\right) - \sum_{c=1}^C \pi_c H(P_c),
\end{equation*}
where $H(P)$ is the Shannon entropy of the probability distribution $P$ given by
\begin{equation}
H(P) =  -\sum_{m=1}^MP(m)\log P(m).
\end{equation}
The $\text{JSD}_{q}$ quantity defined in Eq.~\eqref{eq:JSDq} is in fact the JSD (with a particular weighting) of the maximum likelihood estimates of the $\probvec_c$, so we call $\text{JSD}_{q}$ the \emph{observed JSD}.  This can be shown directly by letting $P_c(m) \to \counts_{c,m}/N_c$ and taking  $\pi_c = N_c/N$ (where $N=\sum_cN_c$),  in the definition of JSD.

The observed JSD is an estimate of the JSD of the underlying probability distributions for circuit $q$.  Even if there is no context dependence, however, each $\text{JSD}_q$ will almost always be non-zero due to ordinary finite-sample fluctuations.
Thus $\text{JSD}_q$ is significantly different from zero only if it is greater than
\begin{equation}
\text{JSD}_{\text{threshold}} = \frac{\lambda_{\text{threshold}}}{2N},
\label{eq:JSDthreshold}
\end{equation}
where $\lambda_{\text{threshold}}$ is the LLR pseudo-threshold of Eq.~\eqref{eq:llrthreshold}. Implicit in this relation is the fact that $\lambda_q$ and $\text{JSD}_q$ are entirely equivalent test statistics.

\subsection{Total variation distance}\label{sec:tvd}
JSD quantifies statistical distinguishability between probability distributions and their average \cite{lin1991divergence}, so an estimate of the underlying JSD is a well-motivated measure of the context dependence of a circuit. However, there are other metrics with other meanings.  One commonly used in quantum information is the total variation distance (TVD) \cite{Verdu2014}. The TVD between two distributions $P_1$ and $P_2$ over $M$ events,  is
\begin{equation}
\text{TVD}(P_1,P_2) = \frac{1}{2}\sum_{m=1}^M|P_1(m)-P_2(m)|.
\end{equation}
The observed TVD for circuit $q$ ($\text{TVD}_q$) is naturally defined by
\begin{equation}
\text{TVD}_q = \frac{1}{2}\sum_{m=1}^M \left| \frac{\counts_{1,m}}{N_1} - \frac{\counts_{2,m}}{N_2} \right|.
\end{equation}
Here the contexts are indexed ``1" and ``2", because the TVD is only defined between two contexts, i.e., when $C=2$.

Even if there is no context dependence, observed TVDs between two contexts are generally non-zero because of finite-sample fluctuations.  It is often useful to correct for this.  Unlike the observed JSD, however, the observed TVD is not simply related to the LLR so there is no simple seudo-threshold for $\text{TVD}_q$. Instead, we introduce the \emph{statistically significant total variation distance} (SSTVD). If statistically significant variation is detected for circuit $q$ using the ICTs, we report $\text{SSTVD}_q = \text{TVD}_q$ for that circuit; when no statistically significant context dependence is detected, the circuit has no SSTVD. That is,
\begin{equation}
\text{SSTVD}_q = \begin{cases}  \text{TVD}_q       & \quad \text{if } \lambda_q > \lambda_{\text{threshold}}, \\
    \text{null}  & \quad \text{else}.
  \end{cases}
\label{eq:SSTVD}
\end{equation}
Note that we do not define $\text{SSTVD}_q$ to be zero when $\lambda_q \leq \lambda_{\text{threshold}}$.  Failure to detect context dependence does \emph{not} imply that this circuit is probably context independent. This is because not rejecting a null hypothesis in a hypothesis test does not imply anything about whether that null hypothesis is true. For example, one or more $\lambda_{q}$ could be just below the pseudo-threshold at a global $5\%$ significance and above the pseudo-threshold at a global significance of $6\%$. Those circuits are therefore quite probably context dependent, meaning that a $\text{SSTVD}_q$ of zero could be misleading.

When analyzing data from many circuits ($Q \gg 1$), it is often useful to summarize any observed context dependence with a single number. One such candidate is the maximum SSTVD over all circuits
\begin{equation}
\max \text{SSTVD}  =  \max_{q} \left[\text{SSTVD}_q\right],
\label{eq:maxSSTVD}
\end{equation}
and we will use this statistic in our examples later. The motivation for $\max \text{SSTVD}$ is that it partially captures worst-case context dependence. For example, without context dependent SPAM, the maximum over gates of the diamond distance between the process matrix for each gate in the two contexts is \emph{lower bounded} by the maximum true TVD over the circuits, divided by the number of gates in the maximizing circuit. The $\max \text{SSTVD}$ is an estimate of this maximal TVD. (This link to diamond distance suggests an interesting alternative to $\max \text{SSTVD}$; $ \max_{q} \left[\text{SSTVD}_q/l(q)\right]$ where $l(q)$ is the length of circuit $q$).  It is also important to note that the value of $\max \text{SSTVD}$ is, in general, strongly dependent on the choice of circuits, even when divided by circuit length, as the most context dependent circuit might not be in the set of circuits chosen.

There are some subtleties to SSTVD, which can become important in slightly unusual circumstances. Perhaps the most significant of these is that the SSTVD of a circuit can \emph{sometimes} significantly over-estimate the true TVD of the circuit. For example, consider a situation whereby the TVD between contexts is the same and fairly small for all circuits, and context dependence is detected in only some of the circuits (because the effect is small, so the chance that it is detected in any particular circuit is low). The circuits in which SSTVD is reported as non-null must have an observed TVD large enough so that the LLR test triggers, and the minimum such observed TVD could be significantly largely than the true TVD. If this is the case, any non-null SSTVD is a significant over-estimate of the true TVD. Subtleties of this sort can be accounted for by looking at additional properties of the observed TVD distribution. However, this is not to suggest that looking at the full observed TVD distribution is always preferable in practice: the SSTVD is a convenient tool for highlighting the rough size of any detected context dependence without requiring subtle, case-specific analysis of a distribution.

\section{Simulated drift detection}\label{sec:sim}
In this section we present a simulated example showing how to use the tools presented above to detect slow drift. This example uses data from GST circuits, but alternatives such as RB circuits could equally have been used. We consider \emph{long-sequence} GST (LSGST) circuits \cite{blume2016certifying} built from two gates: $\pi/2$ rotations around $\sigma_{x}$ and $\sigma_y$.  Each LSGST circuit begins with one of six short state-preparation sequences, followed by one of six short ``germ'' sequences repeated $O(K)$ times, and concludes with one of six short pre-measurement sequences.  These building blocks are chosen so that the collection of LSGST circuits are informationally complete \cite{blume2016certifying,rudinger2017experimental}. Here, $K$ ranges from 0 to 256 with logarithmic spacing, yielding 1405 unique quantum circuits. Below, the size of $K$ is referred to as the ``core'' circuit length. The specific circuits used are given in Appendix~\ref{app:circuits}.

\begin{figure}
\includegraphics[width=1\columnwidth]{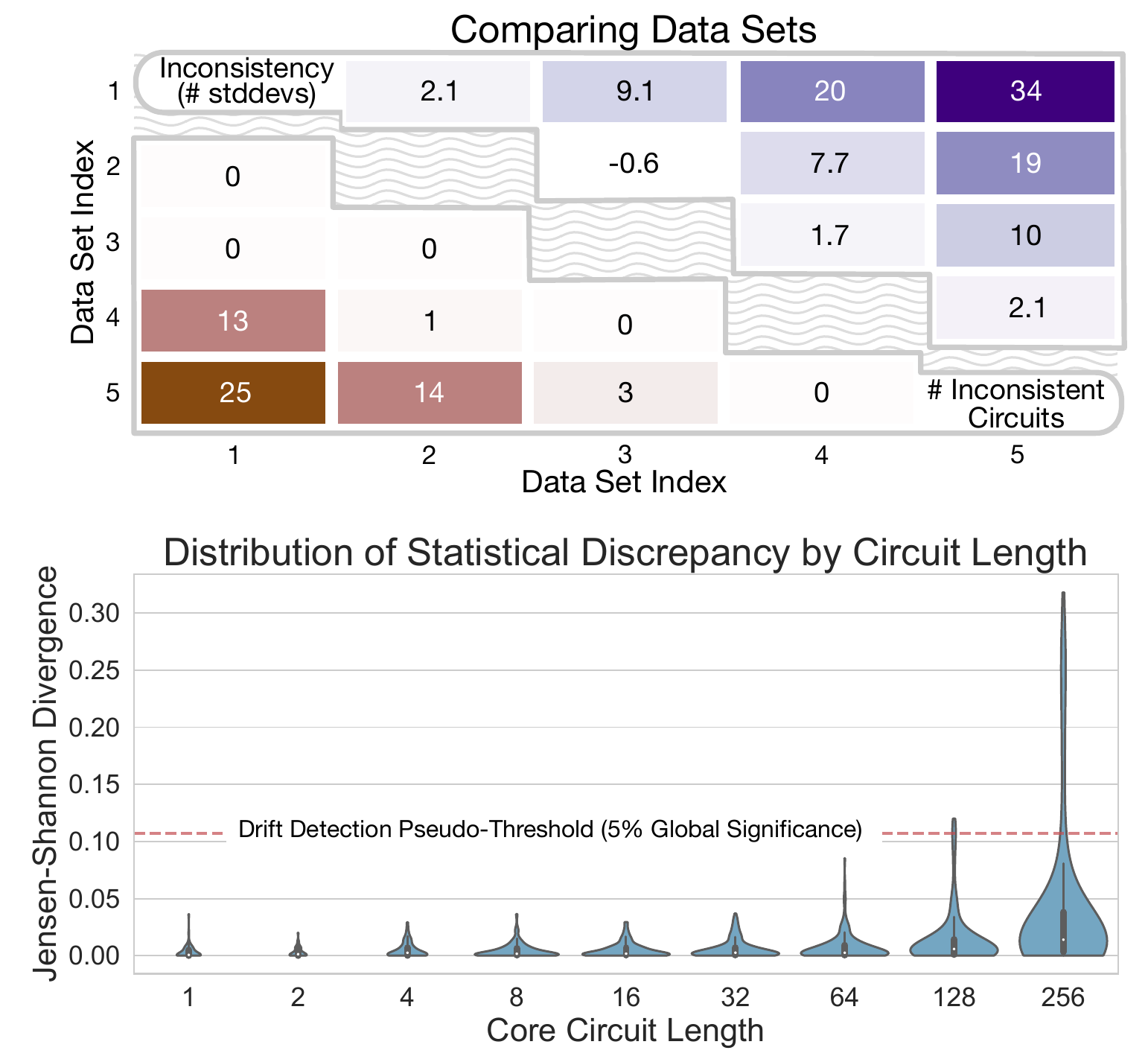}
\caption{\label{fig:heatmaps}An example using our techniques for drift detection on simulated data. Data was obtained by repeating the same 1405 circuits 100 times in each of five time periods.  The circuits contain $\pi/2$ rotations around $\sigma_{x/y}$ and are informationally complete, meaning that they are collectively sensitive to drift in every aspect of gates and SPAM. Drift was modeled as time-dependent over-rotations in both gates, by $(t-1)\cdot10^{-3}$ radians in time period $t=1,2,\dots,5$. Upper plot, upper triangle: $\mathcal{N}_\sigma$ of total model violation for pairwise comparisons between the five pools. Upper plot, lower triangle: the number of circuits that were found to contain statistically significant drift. Lower plot: A violin plot of the estimated Jensen-Shannon divergence (JSD) for each circuit vs.~core circuit length for the $t=1$ to $t=5$ time period comparison (``core'' circuit length is defined in the main text). Any JSD above the pseudo-threshold is significantly non-zero, at $5\%$ global statistical significance, implying that drift has been rigorously detected in the associated circuits. As discussed in the main text, by looking at which circuits have a high JSD it is possible to infer the form of the errors.}
\end{figure}

We simulated repeating these circuits $N=100$ times in each of 5 consecutive time periods $t=1,2,\dots,5$ (the contexts).  In addition to small time-\emph{independent} unitary errors in the gates, we simulated slow drift by adding over-rotations of $(t-1)\cdot10^{-3}$ radians in time periods $t$ to both gates. We tested for drift (context dependence between time periods) using a global significance level of $\alpha = 5\%$. 

There are five contexts (the five time periods), so there are many ways to test for drift: we can implement the tests introduced earlier on all the data (jointly comparing the five contexts) and/or we can implement up to 10 pairwise comparisons between pairs of different time periods (comparing pairs of contexts). We'll demonstrate all of these analyses, resulting in 11 comparisons between contexts in total. Therefore, to guarantee a global significance of $5\%$ we perform each comparison between contexts at a significance of $(5/11) \% \approx 0.45\%$ (this is a Bonferroni correction), with the aggregate LLR test and the ICTs performed for each comparison using the particular multi-test correction procedure specified earlier (so, for example, each aggregate LLR test is performed at $(5/22)\% \approx 0.23\%$ significance).  For the joint comparison of all five time periods, we find that the signed standard deviation of the aggregate LLR $\mathcal{N}_{\sigma}$, defined in Eq.~\eqref{eq:nsigma}, is $\mathcal{N}_{\sigma} \approx 21$;  the threshold for drift detection is only $\mathcal{N}_{\sigma} \approx 2.9$ (as given by Eq.~\eqref{eq:nsigmathreshold} with $\alpha \approx 0.23\%$).  Thus we have detected drift with extremely high confidence. The ICTs test also detects drift, finding 21 circuits to be significant.

To obtain more detailed, diagnostic information, we turn to the pairwise time period comparisons. These results are summarized in Fig. \ref{fig:heatmaps}.  The upper triangle in the upper plot of Fig.~\ref{fig:heatmaps} shows $\mathcal{N}_{\sigma}$ for each pairwise comparison. For the longest time difference comparison $\mathcal{N}_{\sigma} \approx 34$ (the threshold for drift detection is still $\mathcal{N}_{\sigma} \approx 2.9$). The lower triangle in the upper plot of Fig.~\ref{fig:heatmaps} shows the number of circuits that were found to have statistically significant drift for each pairwise comparison. If this is zero \emph{and} the $\mathcal{N}_{\sigma}$ is not statistically significant then drift is not detected for that pairwise comparison; otherwise it is. Therefore, none of the comparisons between neighboring time periods detect drift, but all other comparisons \emph{do} detect drift.  Drift is thus detected whenever the difference in rotation angle between time periods is at least $2\cdot 10^{-3}$ radians. As expected, the statistical significance of the observed effect, as quantified by $\mathcal{N}_{\sigma}$, increased with time delay.  Note that, while no drift was detected between neighboring time periods, we know that drift was present (because we designed the model). This drift could have been made visible to our tools in either of two ways. Firstly, we could have included longer sequences that would be more sensitive to small rotations. Alternatively, we could simply have collected more data.

Fig.~\ref{fig:heatmaps} also demonstrates that these tools allow for a rough diagnosis of the drift, without requiring computationally expensive parameter estimation. The lower plot of Fig.~\ref{fig:heatmaps} shows the distribution of the per-circuit observed JSDs, as defined in Eq.~\eqref{eq:JSDq}, versus ``core'' circuit length (see above), for the longest delay period $t=1$ vs. $t=5$. This shows that the magnitude of the drift grows with circuit length, implying that the gates are drifting, rather than the SPAM. Note that only those circuits with an observed JSD above the pseudo-threshold for statistical significance, given by Eq.~\eqref{eq:JSDthreshold}, have been flagged up by our tests as being context dependent at $5\%$ global significance (there are 25 of them, as shown in the upper plot).  Looking, however, at the trend in the observed JSD distribution versus sequence length also provides additional, if less rigorous \footnote{Note that the correlation of the observed JSD with circuit length could be turned into a test statistic and used in rigorous statistical hypothesis testing.}, evidence of an increase in the underlying JSD with length (without context dependence, the observed JSD would be uncorrelated with circuit length). This highlights the utility of further data analysis, after context dependence has been first detected with statistically rigorous hypothesis testing.

Looking at the specific details of the circuits, we observe that the largest observed JSDs are seen in circuits where the same gate is repeated sequentially many times. This strongly suggests that the gate rotation angles are drifting, rather than the rotation axes (which those circuits would not amplify sensitivity to) or the stochastic error rates (changes in which would manifest in \emph{all} longer sequences). This is, of course, consistent with the simulated error model. Jupyter notebooks that contain this more detailed analysis, and which can be used to repeat and extend these simulations, are included as supplemental material \footnote{See source files for this preprint, listed under ``Other formats''.}.

\section{Experimental drift and crosstalk detection}\label{sec:exp}
\begin{figure}[h!]
\includegraphics[width=1\columnwidth]{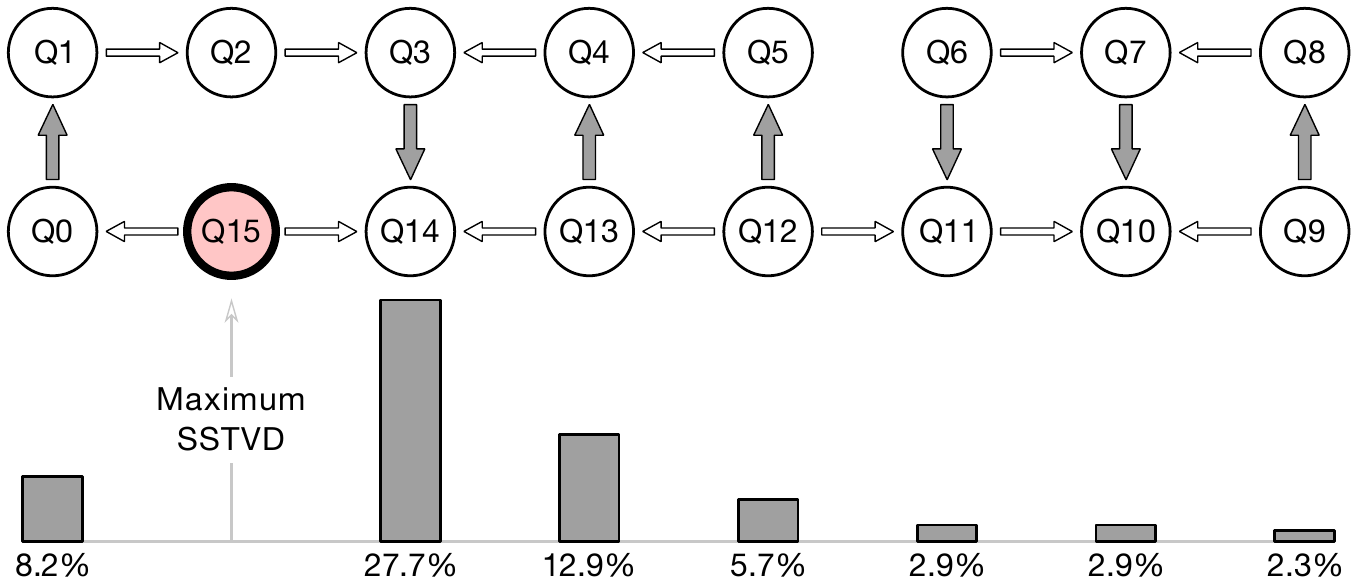}
\caption{\label{fig:ibmqx3results} Quantifying the effect of \cnot gates on the performance of qubit $Q_{15}$ in ibmqx3 \cite{ibmqx}. Top: a schematic of ibmqx3 with $Q_{15}$ highlighted. Circles indicate qubits and arrows denote \cnot gates, pointing from the control to target. Bottom: The effect of driving each of the seven ``ladder-rung'' \cnot gates on short circuits run on qubit $Q_{15}$, as quantified by $\max \text{SSTVD}$, which is an empirical, total-variation-distance based measure that we propose for estimating worst-case context dependence over circuits (see main text). The $\max \text{SSTVD}$ from driving each \cnot is plotted immediately below the corresponding rung in the schematic. The \cnot between qubits $Q_{14}$ and $Q_{3}$ has a large effect on the behavior of circuits on $Q_{15}$, which corresponds to changing the outcome probabilities of a set of short circuits on $Q_{15}$ by 27.7\% in the worst case. The circuits run on $Q_{15}$ were those of linear-inversion gate set tomography, and are discussed in the main text.}
\end{figure}

To further demonstrate the practical utility of our tools, we applied them to detect and quantify drift and crosstalk in the publicly accessible ibmqx3 \footnote{As of June 2018, ibmqx3 has been replaced by ibmqx5, a device similar to ibmqx3.}\cite{ibmqx,qiskit}. This is a 16-qubit superconducting device with connectivity on a 2 $\times$ 8 grid, shown schematically in Fig.~\ref{fig:ibmqx3results}, resembling a ladder. We ran circuits over $\{\mathsf{I}, \mathsf{H}, \mathsf{S}\}$ gates on a single qubit ($Q_{15}$) to see whether:
\begin{itemize}
\item[(I)] The behavior of this qubit was affected by simultaneous \cnot gates applied to various ``rungs'' of the ``ladder''.
\item[(II)] The behavior of this qubit drifted in time.
\end{itemize}

 To do this, we implemented the circuits of \emph{linear inversion} GST (LGST) \cite{blume2013robust} over $\{\mathsf{I}, \mathsf{H}, \mathsf{S}\}$ on $Q_{15}$ in multiple contexts. LGST is the simplest, least experimentally intensive form of GST, requiring only 40 unique circuits for these gates. The exact circuits are listed in Appendix~\ref{app:circuits}, and all the circuits are depth 7 or less.  For each rung, we compare the output of LGST circuits on $Q_{15}$ in the following time-ordered contexts:
\begin{itemize}
\item[(a)] All other qubits idle.
\item[(b)] The \cnot on the rung is applied whenever a gate is applied to $Q_{15}$.
\item[(c)] All other qubits idle.
\end{itemize}
This experimental design was chosen to enable detection and isolation of both drift \emph{and} crosstalk.  If no context dependence is detected between (a) and (c), then we can safely rule out drift. Any context dependence between (a) and (b) may then be ascribed to crosstalk (modulo caveats discussed later).  Access constraints prohibited running all the circuits for a rung in one submission.  Therefore, for each rung, we submitted the circuits for each context [(a) -- (c)] in sequential batches. The delay between executed batches ranged from a few seconds to several minutes, depending on machine availability.

To implement the tests, we picked a global significance of 5\%. To maintain this global significance level, a Bonferroni correction was used to split this 5\% evenly over the comparisons for the seven rungs and the (a) to (b) and (a) to (c) comparisons for each rung (we do not compare (b) to (c) so as to avoid additional local significance dilution). This results in implementing each pairwise context comparison at a significance of $\frac{5}{14}\%$, noting that each pairwise comparison itself contains 40 per-circuit comparisons (the ICTs) and an aggregate comparison, as described earlier.  (The resulting data, along with the full analysis, is provided in supplemental material \footnote{See source files for this preprint, listed under ``Other formats''.})

We detected no drift. That is, for all seven rungs, no change was detected between any (a) and corresponding (c) context. This is interesting in its own right, but it is also critical for the crosstalk detection. This is because it implies that any variation between any (a) and (b) contexts is probably \emph{not} due to random drift -- and thus, if differences are detected, that they are almost certainly due to the \cnot gate on the rung in question.

Our results comparing contexts (a) and (b) for each rung are summarized in Fig.~\ref{fig:ibmqx3results}, where we plot the $\max \text{SSTVD}$ for each rung (see Eq.~\eqref{eq:maxSSTVD}). In all cases, the application of \cnot gates on the other qubit pairs influences the behavior of $Q_{15}$ to a statistically significant degree, as the $\max \text{SSTVD}$ is non-zero (the SSTVD of a circuit is ``null'' if context dependence was not detected for that circuit; see Eq.~\eqref{eq:SSTVD}). The observed maximum SSTVD broadly decreases with the connectivity graph distance between $Q_{15}$ and the driven rung.  Thus closer \cnot gates generally affect $Q_{15}$ more. For the \cnot between Q3 and Q14, one of the two closest rungs to $Q_{15}$, we observed a $\max \text{SSTVD}$ of around $28\%$, corresponding to the gate sequence $\mathsf{HSSSSH}$. For this circuit, out of 1024 measurement results, just 2 ``1'' outcomes were observed in context (a), while 286 ``1'' outcomes were observed in context (b). That is, this suggests that applying the \cnot gate to this rung changed the outcome probabilities of this circuit on $Q_{15}$ by about 28\%.

The obvious cause of changes from contexts (a) to (b) is crosstalk, but there is an important caveat that needs to be addressed before we can conclude this. The circuits on $Q_{15}$ took longer when applying a \cnot to a rung (context (b)) than when implemented in isolation (context (a) or (c)). This is because \cnot gates take substantially longer to implement than 1-qubit gates on ibmqx3 \cite{ibmqx}, and in context (b) a single \cnot was applied in parallel with every gate acting on $Q_{15}$. Thus a change in the output probabilities of $Q_{15}$ from context (a) to (b) could be just due to the circuits taking longer, allowing for more decoherence to build up on $Q_{15}$.

This effect, however, will be independent of the rung being tested, and this allows us to bound this effect. The $\max \text{SSTVD}$s between context (a) and (b) for the three furthest rungs are all approximately equal (see Fig~ \ref{fig:ibmqx3results}), and much lower than the $\max \text{SSTVD}$s for the other rungs. These $\max \text{SSTVD}$s provide a rough baseline for the maximal amount of the context dependence that can be attributed to this timing difference; any excess in the $\max \text{SSTVD}$ above this level is almost certainly due to crosstalk.

To fully isolate the crosstalk caused by a \cnot from any change in circuit performance caused by increased circuit duration, the time for each circuit layer should be fixed for all contexts, which could be more easily incorporated into experiments with lower-level access to a device. This is illustrative of the need to carefully account for all ``nuisance contexts'' that may be unintentionally or unavoidably changing with the context of interest. These nuisance contexts should be removed if possible, or, as here, accounted for when not.

\section{Discussion}\label{sec:srb}
To our knowledge, the tools we have presented and demonstrated herein are the first designed for detecting and characterizing generic context dependence in generic quantum circuits. However, one particular important example of context dependence is crosstalk, and there is already a widely used tool for characterizing crosstalk: simultaneous randomized benchmarking (SRB) \cite{gambetta2012characterization,mckay2017three}. For this reason, we now briefly discuss the relationship between our tools and SRB. In essence, SRB involves comparing a qubit's RB error rates in two contexts, corresponding to (1) leaving neighbor qubits idle, and (2) driving them. This then provides a quantification of crosstalk in terms of the increase in the RB error rate caused by driving neighboring qubits.

Our methods complement those of SRB: our tools are not restricted to RB circuits, but unlike SRB they cannot directly provide a ``crosstalk error rate'' for the gates.  Moreover, our methods can't be applied directly to SRB data, because SRB uses independently sampled (and so almost certainly different) random sequences in each context.  Our methods \emph{can}, however, be used in concert with the SRB analysis if SRB is modified slightly, so that each random sequence appears in both the driven- and undriven-neighbor(s) contexts. With data from circuits of this sort, our tools complement the standard SRB analysis; they provide statistically rigorous crosstalk detection, something not directly addressed by the SRB analysis. Moreover, our tools allows for the testing of each \emph{individual} random SRB sequence for sensitivity to driving, and this can potentially help to identify the main sources of crosstalk (particularly if using varied-sampling-distribution RB methods such as those in Ref.~\cite{proctor2018direct}).

\section{Conclusions}\label{sec:conclusions}
Improving the performance of future quantum processors will require quantifying, understanding, and eventually mitigating a wide variety of context-dependent errors, such as crosstalk \cite{gambetta2012characterization,mckay2017three,proctor2018direct}
and drift \cite{fogarty2015nonexponential}. The techniques presented and demonstrated here are simple, general, and statistically rigorous ways to detect and quantify context-dependent errors, independent of their underlying physical causes. These methods are also computationally lightweight, and can be applied to any collection of quantum circuits on any number of qubits. We therefore recommend that almost all device characterization protocols should be augmented with these tools. They can even be applied to archived data if any context-identifying information, such as time stamps, was kept.  We expect that these techniques will contribute to the toolkit for calibrating and debugging next-generation qubits.  For easy use, they have been integrated into (and documented in) the open-source \texttt{pyGSTi} software package \cite{pygstiversion0.9.4}.  

% ========================= Back Matter ===========================%
\hspace{1 cm}
\section*{Acknowledgements}
We thank Jay Gambetta, Lev Bishop, the IBM Quantum Experience team, and Erik Nielsen for technical assistance.  This paper describes objective technical results and analysis. All statements of fact, subjective views, opinions, or conclusions expressed herein are strictly those of the authors; they do not represent the official views or policies of IBM, IARPA, the ODNI, the Department of Energy, or the U.S. Government. Sandia National Laboratories is a multimission laboratory managed and operated by National Technology \& Engineering Solutions of Sandia, LLC, a wholly owned subsidiary of Honeywell International Inc., for the U.S. Department of Energy's National Nuclear Security Administration under contract DE-NA0003525. This material was funded in part by the U.S. Department of Energy, Office of Science, Office of Advanced Scientific Computing Research, and also by the Office of the Director of National Intelligence (ODNI), Intelligence Advanced Research Projects Activity (IARPA).
\bibliography{Bibliography}

\appendix

\section{Circuit details}
\label{app:circuits}
 In this appendix we describe the sets of quantum circuits used in the simulations and experiments of the main text. The circuits are from two forms of gate set tomography (GST) \cite{blume2016certifying,merkel2013self,greenbaum2015introduction,dehollain2016optimization}: Long-sequence GST (LSGST) \cite{blume2016certifying}  circuits are used for the simulations, while linear-inversion GST (LGST) \cite{blume2013robust} circuits are used for the experiments on ibmqx3. Below we only specify the circuits used, not how this set of circuits were chosen. For more information on how to choose GST circuits, see Ref.~\cite{blume2016certifying} and the Jupyter notebooks accompanying this paper.

Following the notation of Ref.~\cite{blume2016certifying}, the idle gate and gates corresponding to $\pi/2$ rotations around $\sigma_{x}$ and $\sigma_y$ are denoted by $G_i$, $G_x$, and $G_y$, respectively. The Hadamard and phase gates are denoted by $G_h$ and $G_s$, respectively, where the phase gate is the unitary that maps $\ket{x} \to i^{x}\ket{x}$ for $x=0,1$. The null gate operation of ``do nothing for no time'' is denoted by ``$\{\}$''.  Circuits are specified in operation order, \emph{not} matrix multiplication order.  For example, the sequence denoted $G_hG_s$ means ``perform a Hadamard gate, followed by a phase gate".

To succinctly list the circuits used in the simulations and experiments, it is necessary to first review the structure of GST circuits. Although not necessary, the GST circuits herein fix all state preparations to the $\ket{0}$ state, and all measurements to be in the $\sigma_z$ basis, so we'll specialize to that case. All GST circuits contain one of several short gate sequences at the beginning of the circuit, as well as another sequence at the end. This is to achieve tomographic completeness, by simulating informationally
complete state preparations and measurements.  These short sequences are referred to as \emph{fiducials}. Given a gate set $\mathcal{G}$, a set of preparation fiducials $\mathcal{F}^{(p)}$ and a set of measurement fiducials $\mathcal{F}^{(m)}$, the collection of LGST circuits is the set of all circuits of the form:
\begin{align*}
F, & \quad \forall F \in\mathcal{F}^{(p)}\cup\mathcal{F}^{(m)},\\
F_pF_m, &\quad\forall F_p\in\mathcal{F}^{(p)},\quad \forall F_m\in\mathcal{F}^{(m)},\\
F_pGF_m, &\quad\forall F_p\in\mathcal{F}^{(p)},\quad \forall G\in\mathcal{G},\quad \forall F_m\in\mathcal{F}^{(m)}.
\end{align*}
Note that some circuits may appear more than once when iterating over all three forms of circuit and all possible combinations of gates, preparation fiducials and measurement fiducials. (And, naturally, a circuit is only added to the list of LGST circuits once). From above, it follows that to define a set of LGST circuits it is only necessary to specify the sets $\mathcal{G}$, $\mathcal{F}^{(p)}$ and $\mathcal{F}^{(m)}$. For the experiments run on ibmqx3, we used the circuits of LGST with:
\begin{align*}
\mathcal{G}&=\{G_i,G_h,G_s\},\\
\mathcal{F}^{(p)}&=\{\{\}, G_h, G_hG_s, G_hG_sG_s\},\\
\mathcal{F}^{(m)}&=\{\{\}, G_h, G_sG_h, G_hG_sG_h\}.
\end{align*}

In addition to the circuits of LGST, LSGST uses a further collection of sequences constructed from powers of a set of \emph{germs}. Like the preparation and measurement fiducials, the germs are short sequences of gates from $\mathcal{G}$. Denote the germ set by $
\mathbb{G}$, with the length of germ $g$ denoted by $\ell(g)$. For LSGST we also need to choose a maximum ``germ power'' $L_{\max}=2^k$ for some positive integer k.  LSGST consists of all the circuits of LGST along with all gate sequences of the form
\begin{align*}
F_pg^{\left\lfloor\frac{L}{\ell(g)}\right\rfloor}F_m, \quad\forall g\in\mathbb{G}, \quad\forall L\in\{1,2,4,\ldots, L_{\max}\},
\end{align*}
where, as above, $F_p$ and $F_m$ run over all preparation and measurement fiducials, respectively. Again, these circuits may not all be unique, or unique from the set of LGST circuits that they are combined with.

For the simulations presented in the main text to illustrate drift detection, we used LSGST circuits with $L_{\max}=256$ and:
\begin{align*}
\mathcal{G}&=\{G_x,G_y\},\\
\mathcal{F}^{(p)}&=\mathcal{F}^{(m)}=\{\{\},G_x,G_y,G_x^2,G_x^3,G_y^3\},\\
\mathbb{G}&=\{G_x, G_y, G_xG_y, G_x^2G_y, G_xG_y^2, G_x^2G_yG_xG_y^2\}.
\end{align*}
This results in 1405 circuits, as stated in the main text.
\end{document}